\documentclass[prb,twocolumn,superscriptaddress,floatfix]{revtex4}
\usepackage[]{graphicx}
\usepackage[]{amsmath}
\usepackage[]{bm}  

\renewcommand{\vec}[1]{\protect\bm{#1}}
\newcommand{\crs}{\protect\bm{\times}}

\newcommand{\dif}{\mathrm{d}}

\newcommand{\etal}{{\it et al.\/}}
\newcommand{\F}{\mathcal{F}}

\begin{document}
\bibliographystyle{apsrev}

 
\title{The role of electron scattering in magnetization relaxation in 
  thin Ni$_{81}$Fe$_{19}$ films}
\author{S.~Ingvarsson}
\affiliation{IBM Research Division, T.~J.~Watson Research Center, Yorktown Heights, NY 10598}
\author{L.~Ritchie}
\author{X.~Y.~Liu}
\author{Gang Xiao}
\affiliation{Physics Department, Brown University, Providence, RI 02912}
\author{J.~C.~Slonczewski}
\author{P.~L.~Trouilloud}
\author{R.~H.~Koch}
\affiliation{IBM Research Division, T.~J.~Watson Research Center, Yorktown Heights, NY 10598}

\date{\today}

\begin{abstract}
We observe a strong correlation between magnetization relaxation and electrical resistivity
in thin Permalloy (Ni$_{81}$Fe$_{19}$, ``Py'') films.  Electron scattering rates in the films
were affected by varying film thickness and deposition conditions.  This shows that the
magnetization relaxation mechanism is analogous to ``bulk'' relaxation, where phonon
scattering in bulk is replaced by surface and defect scattering in thin films. Another
interesting finding is the increased magnetization damping with Pt layers adjacent to the Py
films. This is attributed to the strong spin-orbit coupling in Pt, resulting in spin-flip
scattering of electrons that enter from the Py.
\end{abstract}


\maketitle


\section{Introduction}
The Gilbert form of the Landau-Lifshitz equation describes the small angle precession of
magnetization in a ferromagnet,
\begin{equation}
    \label{eq:LLG}
    \frac{\dif\vec{M}}{\dif t} = -\gamma \vec{M} \crs \vec{H}_{\text{eff}} -
    \frac{\alpha}{M} \vec{M}\crs\frac{\dif \vec{M}}{\dif t}~~.
\end{equation}
Here $\vec{M}$ is magnetization and $\gamma = g\left| e\right|/2mc$ is the gyromagnetic
ratio, and $\alpha$ is the Gilbert damping coefficient that affects the magnitude of the
viscous damping term. $\vec{H_{\text{eff}}}$ is the effective magnetic field seen by the
magnetization, and is expressed in terms of the free energy as $\vec{H_{\text{eff}}} =
-\nabla_{\vec{M}}\F$.
The Gilbert damping coefficient $\alpha$, controls how rapidly the magnetization equilibrates
in the absence of external stimulus.  This obviously makes $\alpha$ a key parameter in the
description of high speed dynamics in magnetic materials.  A few areas where $\alpha$ plays a
vital role are in devices that rely on fast magnetization reversal (e.g.\ in Giant
Magnetoresistive-~\cite{RussekJAP00}, Magnetic Tunnel Junction-~\cite{KochPRL98} or other
spintronic devices), current induced magnetization reversal~\cite{KatinePRL00}, generation
of microwaves by spin currents~\cite{TsoiPRL98} etc.

By the 1970s it had been shown that intrinsic magnetization relaxation in transition metal
ferromagnets could be explained by electron scattering by phonons and magnons.  The former
process, mediated by the spin-orbit interaction, occurs both
with~\cite{KamberskyCJPB76,KamberskyCJP70}, and
without~\cite{KamberskyCJP70,KorenmanPRB72,KorenmanPRB74,BergerJPCS77} the accompaniment of a
spin-flip. In the former case $\alpha \sim \tau^{-1}$, where $\tau^{-1}$ is the electron
scattering rate.  In the latter case the angular momentum relaxes as scattered electrons
repopulate the magnetization-direction dependent Fermi volume, and $\alpha \sim \tau$ is
expected.  Magnon modes can relax through exchange interaction with a conduction electron,
causing its spin to flip (this can be viewed simplistically as s--d exchange accompanied by a
spin-flip of the s-electron)~\cite{HeinrichPSS67,MitchellPR57,KittelPR56}. The conduction
electron spin then relaxes to the lattice through the spin-orbit interaction. This also
results in $\alpha \sim \tau^{-1}$ to leading order.

More recently, new effects were predicted for ultrathin films and for multilayers that would
contribute to the effective $\alpha$ in Eq.~(\ref{eq:LLG}). Arias \etal\ showed that under
certain conditions in ultrathin films the uniform mode ($k=0$), excited by ferromagnetic
resonance (FMR), can be scattered into $k\neq 0$ magnons by anisotropic surface defects,
where $k$ is the wavevector~\cite{AriasPRB99}. This two-magnon scattering theory was used by
Azevedo~\etal\ to explain experimental results of FMR linewidth and resonance field in NiFe
films~\cite{AzevedoPRB00}. Berger predicted that transfer of electron spin angular momentum
between two ferromagnetic layers, separated by a nonmagnetic layer, contributes to the
magnetization relaxation (i.e.\ $\alpha$)~\cite{BergerPRB96}.  The experimental results of
Urban~\etal\ confirmed an increase in FMR-linewidth with two layers of Fe separated by a
nonmagnetic layer, compared to a single layer of Fe~\cite{UrbanPRL01}.

We have studied FMR and electronic transport in NiFe-films (Ni$_{81}$Fe$_{19}$, ``Py'' for
short), in which the electronic scattering rates were affected over a wide range by: (a)
changing the surface scattering contribution by varying the film thickness, (b) changing film
deposition conditions, and (c) choosing different interfaces and surface treatment.  We
observe a strong correlation between the Gilbert damping coefficient and resistivity, i.e.\
$\alpha\propto\rho$ in our single layer Py-samples.  Our results show that the dominant
magnetization relaxation mechanism in these samples involves electron scattering, and is
seemingly insensitive to whether the scattering occurs within the ``bulk'' of the films or at
the surface.  This explains why $\alpha$ is observed to increase with decreasing film
thickness~\cite{PlatowPRB98}.  It also implies that the effective $\alpha$ in magnetic
devices, small in at least one dimension, made with transition metals or alloys, is expected
to be considerably larger than that intrinsic to the bulk material, due to an increased
surface/volume ratio and to enhanced spin relaxation at interfaces~\cite{ZhangPRL96}.  Our
data also appear inconsistent with the two-magnon scattering theory.  Further, we observe
that with nonmagnetic (NM) Pt enclosing our Py-films, the magnetization relaxation is
significantly enhanced, in addition to the electron scattering related mechanism above.  The
enhancement is attributed to spin relaxation of conduction electrons that leave the Py-layer,
both at the interfaces and within the NM-layers.


\section{Experiment}
Our NiFe films were deposited by dc-magnetron sputtering in a vacuum system with a base
pressure of $2\times 10^{-8}$~torr. During deposition they were (all but one series) exposed
to a uniform magnetic field of $\sim 150$~Oe to induce uniaxial in-plane anisotropy. X-ray
results show that the Py films are (111) textured.  We made two series of samples of
Si/SiO$_2$/Py/PR, where PR is photoresist, used to protect the film from oxidation. One of
these (called o-Py) was grown in a uniform applied magnetic field, the other (called d-Py)
without a deliberately applied field. Both series have a clearly defined easy axis, the
direction of which in the d-Py was presumably defined by the Earth's field. Resistivity
measurements on these samples were made using the van der Pauw method.  They indicate that
the d-Py series has more disorder than the o-Py (``ordered''). We also studied the effect of
adjoining Py layers with 80~\AA\ thick NM-metallic layers, i.e.\ Si/SiO$_2$/X/Py/X, where X
is Cu, Nb, or Pt. The Py thickness within each sample series was varied by depositing a
terraced structure on a single wafer using a movable shadow mask. The films were then
lithographically patterned into arrays of discs, 1 or 2~mm in diameter. From each deposition
we thus obtained a series of samples of different thickness, with minimal variations in
growth conditions. We also made one sample of Si/SiO$_2$/Py, 1000~\AA\ thick, and ion-milled
it several times, measuring its thickness and magnetic properties between millings.

To obtain the Gilbert damping in our samples we measured their in-plane magnetic
susceptibility in an FMR-experiment with swept frequency and fixed dc magnetic field.  The
experimental setup is essentially the same as that of
Korenivski~\etal~\cite{KorenivskiIEEETM96}. The magnetic softness of Py ($H_c\leq 4$~Oe)
allows the experiments to be done with applied dc-fields $H \leq 150$~Oe.  Our films are
thinner than the skin depth at the corresponding resonance frequencies, i.e.\
$\omega_r/(2\pi)\leq 3.5$~GHz for the uniform mode of spin precession.  The exchange
stiffness in Py causes higher order spin wave modes to appear at much higher frequencies.
The ac-field is considered uniform throughout the films and a quasistatic approximation
relates the internal and external fields.  These assumptions are supported by the Lorentzian
lineshape of our resonance peaks, and holds even for the thickest samples of $\sim 1000$~\AA.
The FMR experiments were done at a small precession-cone angle, the ratio of the amplitudes
of the ac- and dc-fields being $\sim 10^{-4}$.  There was no detectable change in the FMR
(susceptibility) when the rf-power was increased by 15~dB.
 
\begin{figure}[ht]
 \begin{center}
 \includegraphics[width=3.375in]{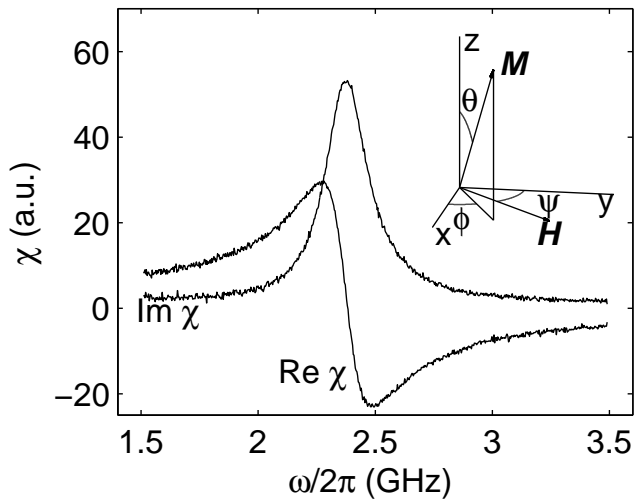}
 \end{center}
 \caption{Real and imaginary part of susceptibility as a function of frequency of a 80Nb/477Py/80Nb
sandwich structure (numbers denote layer thickness in \AA) structure with $H_{dc}=60$~Oe along
the easy axis of the film, $\phi =0$, and $\theta = \psi = \pi /2$.  Also shown, the coordinate
system used to describe the free energy of the film lying in the $x-y$-plane with the easy axis
in the $x$-direction.}
 \label{fig:coordsystem}
\end{figure}

\begin{figure}[ht]
 \begin{center}
 \includegraphics[width=3.375in]{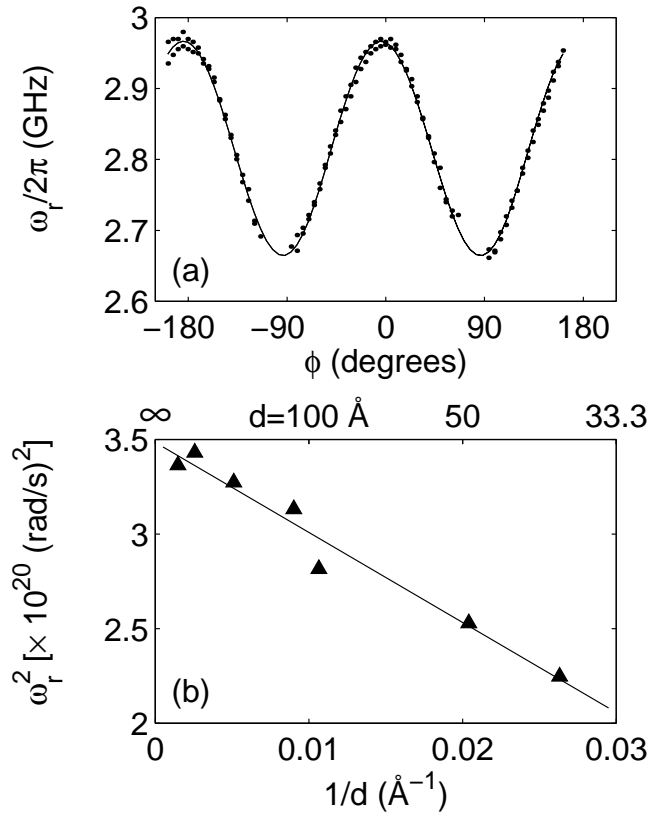}
 \end{center}
 \caption{(a) The dependence of the resonance frequency on the in-plane magnetization angle
   in a 682~\AA\ thick sample from the o-Py series, with $H_{\text{dc}} = 90$~Oe in the plane
   of the film. From a fit to the oscillatory angular dependence we obtain the in-plane
   anisotropy field, here $H_k = 10$~Oe.
   (b) Resonance frequency $\omega_r$ versus sample thickness $d$
   for the o-Py series.  The shift in $\omega_r^2$ scales as $1/d$. This is accounted for by
   surface anisotropy, in accord with Eq.~\ref{eq:resinplane}.  The line is a least squares fit
   and corresponds to $K_s = 0.28$~erg/cm$^2$.} \label{fig:wresfitandwres}
\end{figure}

The conditions above allow us to fit the susceptibility very well with a linearized form of
Eq.~(\ref{eq:LLG}).  The free energy $\F$, includes the Zeeman energy, a demagnetization
term, uniaxial in-plane anisotropy and a uniaxial out-of-plane (interface) anisotropy term.
In the coordinate system shown in Fig.~\ref{fig:coordsystem}, under the assumption that the
applied dc magnetic field $\vec{H}$ is in-plane, $\F$ can be expressed as,
\begin{widetext}
\begin{equation}
    \label{eq:Finplane}
    \F = - MH\sin\theta \sin(\phi + \psi ) +
                      2\pi M^2\cos^2\theta 
	        + K_u\sin^2\theta\cos^2\phi +
	              2\frac{K_s}{d}\cos^2\theta
\end{equation}
\end{widetext}
where $K_u$ is an in-plane uniaxial anisotropy constant, $K_s = \left( K_{s1}+K_{s2}\right)
/2$ is a surface anisotropy constant representing the average anisotropy of the upper and
lower surfaces, and $d$ is the film thickness. The in-plane uniaxial anisotropy is determined
by fitting the angular variation of the resonance frequency, $\omega_r$, in the plane of the
samples.  Typical results are shown in Fig.~\ref{fig:wresfitandwres}, which displays a fit to
$\omega_r/2\pi$ as a function of the equilibrium magnetization angle, $\phi$. 

For the free energy given in Eq.~(\ref{eq:Finplane}) the resonance
frequency is given by,
\begin{widetext}
\begin{equation}
    \label{eq:resinplane}
    \left(\frac{\omega_r}{\gamma}\right)^2 = 
     \left(H\sin \left(\phi + \psi \right) + 
	4\pi M -
	\frac{2K_u}{M}\cos^2\phi +
	\frac{4K_s}{dM}\right) 
    \times\left(H\sin \left(\phi + \psi \right) - 
	\frac{2K_u}{M}\cos \left(2\phi \right)\right)~~.
\end{equation}
\end{widetext}
With the applied field and the magnetization coincident with the easy axis of the sample,
i.e.\ $\phi = 0$, $\psi=\theta=\pi /2$, for each series of samples we use the leading terms
in the field dependence of $\omega_r$ in Eq.~(\ref{eq:resinplane}) to get an initial estimate
of $\gamma$.  Similarly we use the thickness dependence of $\omega_r$ to obtain an initial
value for $K_s$, as shown in Fig.~\ref{fig:wresfitandwres}~(b). Subsequently we obtain the
Gilbert damping parameter $\alpha$, and refined values for $K_s$ and $\gamma$ by fitting the
complex susceptibility. With this approach we arrive at a self-consistent result where a
constant $K_s$ accounts for the shift in the resonance frequency as a function of film
thickness \emph{and} where the $g$-value is constant for any given series of samples.

Even without our careful determination of the abovementioned values we could get a good
estimate of $\alpha$, since its effect on the resonance curve can not be mimicked by
adjusting other parameters. However, as damping increases, the resonance peak gradually
disappears and it becomes difficult to estimate $\alpha$ accurately, even with our fitting
procedure.  This is reflected in the size of the errorbars in Fig.~\ref{fig:alpharho}~(c). We
emphasize that in our treatment $\alpha$ is the dimensionless Gilbert damping coefficient and
not the frequency linewidth.  Although we observe the linewidth following the same trends as
$\alpha$ as a function of film thickness, we believe that the fitting procedure gives more
reliable results than measuring the linewidth. Also, $\alpha$ represents the total effective
damping, including both intrinsic and extrinsic damping, such as surface damping.


\section{Results and discussion}
\begin{figure}[ht]
 \begin{center}
     \includegraphics[width=3.0in]{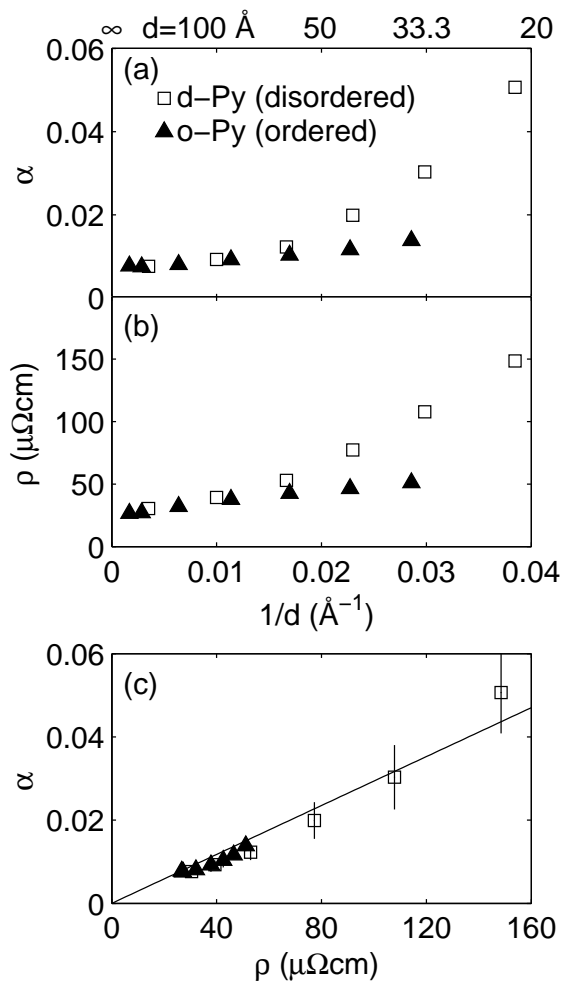}
 \end{center}
 \caption{(a) Gilbert damping coefficient $\alpha$, and (b) resistivity $\rho$ of the d-Py
   (disordered) and o-Py (ordered) series as function of inverse film thickness $1/d$, at
   room temperature.  The two series exhibit quite different thickness dependence. (c)
   displays the correlation between $\alpha$ and $\rho$.} \label{fig:alpharho}
\end{figure}

The Gilbert damping coefficients $\alpha$, for the two PR-coated series (o-Py and d-Py) are
displayed in Fig.~\ref{fig:alpharho}~(a), as function of inverse film thickness $1/d$.
Values for the thickest films approach the bulk damping, but as the thickness decreases
$\alpha$ increases dramatically.  In the o-Py the damping doubles from the thickest film to
the thinnest.  The effect is much more pronounced in the d-Py, where the thinnest film has
roughly six-fold the bulk damping value. The room temperature resistivity
$\rho\left(d\right)$, of these two series is shown in Fig.~\ref{fig:alpharho}~(b).  The
changes in $\rho$ reflect quite accurately the corresponding changes in $\alpha$, i.e.\
$\Delta\rho/\rho_0 \simeq \Delta\alpha/\alpha_0$, where the subscript 0 refers to thick film
values. This suggests the existence of a simple relationship between $\alpha$ and $\rho$,
which is addressed below. The reason for plotting these data as function of $1/d$ is, that if
one naively assumes that $\alpha$ can be separated into independent bulk and surface
contributions a $1/d$-dependence is expected as the surface/volume ratio changes. This is
analogous to the assumption made for electron scattering in the Fuchs-Sondheimer theory of
surface scattering~\cite{Tellier82}.

However, neither the o-Py nor the d-Py series exhibit a perfectly linear relationship between
$\alpha$, or $\rho$, and $1/d$, although the deviation of the o-Py samples from a straight
line appears small. Neglecting the small deviation, a Fuchs-Sondheimer-type analysis of the
o-Py resistivity results in a bulk resistivity at room temperature of $\rho_{\text{b}} =
24$~$\mu\Omega$cm and mean-free-path $\lambda = 96$~\AA\ (the corresponding low temperature,
or residual, values are $\rho_{\text{b,res}} = 14$~$\mu\Omega$cm and $\lambda_{\text{res}} =
215$~\AA). It is clear from the significant departure of $\rho$, in the d-Py series, from
linear $1/d$-dependence, that the simple assumptions of the Fuchs-Sondheimer model do not
hold there. Additional scattering mechanisms such as impurity and grain boundary scattering
should be taken into account~\cite{Tellier82}. In any case, it is evident from the small
resistivity ratios in both series, $\rho_{295 \text{K}}/\rho_{\text{res}}$ ranging from 1.85
for the thickest films to 1.1 for the thinnest films), that scattering associated with film
surfaces and defects (including grain boundaries) accounts for the major share of the total
resistivity in all of these films.  The strongest $T$-dependent contribution to $\rho(T)$ at
low $T$ is proportional to $T^2$, which is attributed to electron-electron and
electron-magnon scattering~\cite{CampbellFertinWohlfarth}.

When the data in Figs.~\ref{fig:alpharho}~(a) and (b) are plotted together, as in
Fig.~\ref{fig:alpharho}~(c), they fall on a single curve. Note that the thickness of the
films in the overlap region is quite different for the two series.  For instance, the
thinnest o-Py film in Fig.~\ref{fig:alpharho}~(c) is 35~\AA\ thick, and it corresponds
roughly to a d-Py film of thickness 65~\AA.  At least to leading order, a simple
proportionality, describable by \emph{one} constant, of experimental $\alpha$ to $\rho$ as
functions of $d$ is apparent in Fig.~\ref{fig:alpharho}~(c). However, examination of the data
in Fig.~\ref{fig:alpharho}~(a) and~(b) reveals that even \emph{two} parameter fits of
$\alpha$, or $\rho = a + bd^{-n}$ to the data would not be fully satisfying.  These remarks
suggest immediately that a significant contribution to viscous damping in very thin
ferromagnetic metals is connected with the electron scattering giving rise to resistivity in
a general way not depending on whether the electron scattering is by phonons, defects, or
surface irregularities.


Theoretical estimates by Kambersky~\cite{KamberskyCJPB76} are particularly helpful in
distinguishing a low-temperature damping term proportional to the electron scattering time
$\tau$ from a high-temperature term proportional to $\tau^{-1}$, when electron scattering by
phonons is prevalent. Electron scattering in our samples is dominated by surface and defect
scattering.  It is mainly caused by fixed electrostatic potentials associated with
compositional interdiffusion and structural irregularities.  It is therefore more appropriate
to use the results of Heinrich~\etal~\cite{HeinrichPSS67} to account for our data.  They
considered the effect of s-electron spin relaxation on s--d exchange and magnetization
relaxation:

By definition, the equation $\alpha=\lambda/\gamma M$ relates the dimensionless Gilbert
damping coefficient to the Landau-Lifshitz parameter $\lambda$.  Combination of Eq.~(21) in
Ref.~\onlinecite{HeinrichPSS67} and the assumption that the spin-relaxation rate
$\tau_s^{-1}\sim\zeta\tau^{-1}$, where $\zeta$ is a constant, and $\tau^{-1}$ is the ordinary
electron scattering rate (characteristic of electrical resistivity, i.e.\ it includes both
scattering events that are accompanied by a spin-flip and those that are not), yields the
estimate,
\begin{equation}
    \alpha\simeq\frac{\zeta\gamma m^* k_{\text{F}}}{2\pi^{2} M \tau}~~.
\end{equation}
Here $\gamma$ is the free-electron gyromagnetic ratio, $m^*$ the effective mass of the
s-electron, $k_{\text{F}}$ the Fermi wavevector.  We have neglected a term $H/\left( 4\pi
M\right) \sim 10^{-2}$ compared to unity. We now eliminate the scattering rate with the Drude
conductivity formula $\sigma = \rho^{-1} = ne^2 \tau/m^*$.  Further, we crudely estimate
$M\simeq\mu_{\text{B}} n$ with $\mu_{\text{B}}=\hbar\gamma/2$ the Bohr magneton, and $n$ the
atomic density, to obtain,
\begin{equation}
    \frac{\alpha}{\zeta\rho}
         \simeq\frac{e^2 k_{\text{F}}}{\pi^2 \hbar}
	 \simeq 3.7\times 10^5 (\Omega\text{m})^{-1}~~,
    \label{eq:alpharho}
\end{equation}
estimated for a 3d metal or alloy.

The solid line in Fig.~\ref{fig:alpharho}~(c), obtained by a linear least squares fit
constrained to go through the origin, has a slope of $3\times 10^4$~($\Omega$m).  To satisfy
Eq.~(\ref{eq:alpharho}) we must then have $\zeta\sim 10^{-1}$, i.e.\ $\tau_s^{-1} \sim
10^{-1}\tau^{-1}$.  For comparison we observe that the diffuse scattering of an electron is
represented by a random walk process, where the spin diffusion length $l_{\text{sf}} =
\lambda\left( \tau_s /\tau \right)^{1/2}$, $\lambda$ is the mean free path of the electron,
and $\tau_s/\tau$ is the number of spin-preserving scattering events before the spin is
flipped.  With published values for Py of $l_{\text{sf}}=55$~\AA~\cite{SteenwykJMMM97}, and
mean free paths $\lambda_{\downarrow} \alt 6$~\AA\ and $\lambda_{\uparrow} =
46$~\AA~\cite{GurneyPRL93}, we estimate the ratio $\left( \tau_s /\tau\right)^{-1}$ to be
0.01 and 0.7 for the down and up spin bands, respectively.  Our result lies well within that
range.


Alternative to the electron scattering mechanism, one may also consider that of two-magnon
scattering~\cite{AriasPRB99}. Since both mechanisms involve interfacial effects, they both
predict increased damping with decreasing film thickness.  But, to account for the plot in
Fig.~\ref{fig:alpharho}~(c) one must show how two-magnon scattering predicts the \emph{equal
damping} in two films having \emph{unequal thicknesses} yet exhibiting \emph{equal
resistivities} because of a compensating difference in their preparation. The surface
roughness of our samples, determined by atomic force microscopy (AFM), was the same $6\pm
1$~\AA$_{\text{rms}}$, both within \emph{and} between the two series.  Furthermore, the
measured surface anisotropies of the o-Py and the d-Py samples are constant and identical,
i.e.\ $K_s = 0.28$~erg/cm$^2$ in both cases.  It therefore appears improbable that (a) the
increase in $\alpha$ with decreasing film thickness is caused by variations in surface
anisotropy, and (b) the similarity with the changes in $\alpha$ and $\rho$ is simply a
coincidence. Also the theory predicts a field dependence of $\alpha$, i.e.\ a decrease as the
applied field is reduced since the two-magnon contribution vanishes in the limit of zero
field~\cite{AriasPRB99}.  We observe a constant $\alpha$ in the field range 150 down to
30~Oe, at which point $\alpha$ starts to \emph{increase} due to incomplete saturation of the
films.

\begin{figure}[ht]
 \begin{center}
     \includegraphics[width=3.375in]{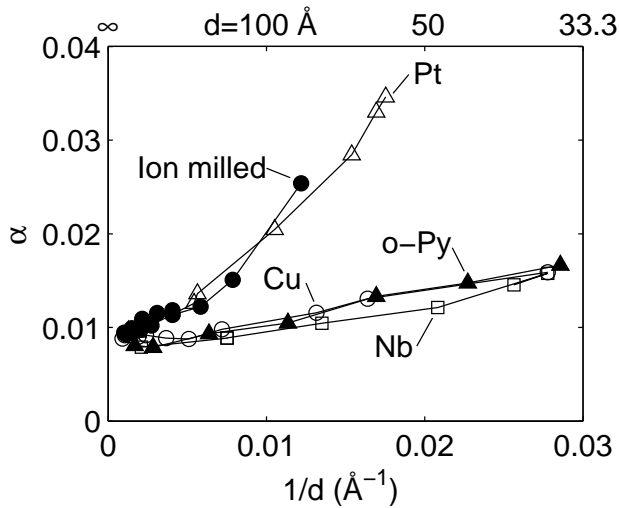}
 \end{center}
 \caption{Gilbert damping coefficient $\alpha$, of Py adjoined with Cu, Nb, and Pt. As a
   reference we plot the o-Py series. Also shown is a Py-film that was ion milled from one
   side.} \label{fig:allalpha}
\end{figure}

The effect of adjoining Py-layers with different NM-metallic layers on the magnetization
damping is shown in Fig.~\ref{fig:allalpha}, with the o-Py series as a reference.
For thick samples the damping parameter remains constant, equal to the bulk value.  However,
as the film thickness decreases $\alpha$ increases rapidly, the rate of increase depending on
the type of interface. In the Cu-, Nb- and o-Py series the damping at
$d_{\text{Py}}\sim35$~\AA\ is approximately twice the bulk value in those samples.  It is
apparent that the effect on $\alpha$ of adjoining the Py films with Cu or Nb is the same as
protecting the Py with photoresist, i.e.\ the fact that these thin \emph{metallic} films meet
with Py is insignificant in these cases. In contrast a very pronounced effect was observed
with Pt-coated surfaces. Assuming a linear dependence of $\alpha$ on $1/d$ the slope of the
Pt-coated material is more than 4.5 times the slope for the other overlayers, and the
$\alpha$-value at $d_{\text{Py}}=57$~\AA\ is almost 4 times the bulk value.  This effect was
observed independently by Mizukami~\etal~\cite{MizukamiJJAP01}.  We have also found that
successive thinning of a Py-film by ion milling led to greatly enhanced damping. As can be
seen in Fig.~\ref{fig:allalpha} the effect of ion milling one side of the sample is at least
equivalent to that of having two Py/Pt interfaces.

The great increase in $\alpha$ in the ion-milled sample most likely arises from increased
electron scattering at the surface due to surface damage caused by the ion-milling process.
Again we observe a constant $\alpha$ as a function of applied field, arguing against
two-magnon scattering effects similar to those observed by LeCraw~\etal~\cite{LeCrawPR58} in
single crystal garnets. The Cu-coated samples had a surface roughness of $\left( 25\pm
3\right.$~\AA$\left._{\text{rms}}\right) $. The Pt-coated samples were a factor of 3-4
smoother $\left( 7\pm 2\right.$~\AA$\left._{\text{rms}}\right)$.  Despite having much rougher
interfaces than both the o-Py and the Pt-coated samples, the Cu-coated samples did not show
enhanced damping over the uncoated o-Py and much less damping than the Pt-coated samples, as
can be seen in Fig.~\ref{fig:allalpha}. These observations confirm that the increased
$\alpha$ in the Pt-coated series is not caused by surface roughness effects and point towards
either an interface scattering effect or to some intrinsic property of the capping layers, or
both. Berger~\cite{Bergerprivate} has predicted, for nonmagnetic layers in contact with a
magnetic layer, a contribution to $\alpha$ from exchange coupling between localized magnetic
spins and conduction electrons accompanied by spin flip scattering both at interfaces
\emph{and} from interaction with phonons in the nonmagnetic layers through the spin-orbit
interaction. The latter would imply that layers with strong spin-orbit interaction, such as
Pt, would provide a more effective damping than e.g.\ Cu, in qualitative agreement with our
results.  Recently, a paper by Tserkovnyak~\etal~\cite{TserkovnyakPRL02} caught our
attention.  They presented a model of this same system based on the idea that the
magnetization precession in the FM layer drives a spin current into the NM layer, where any
spin imbalance is assumed to relax.  Thus the same applies in their case, that Pt should
provide a large enhancement due to its strong spin-orbit coupling and effective spin
relaxation.

We did not study the resistivity of the trilayer films coated with Cu, Nb and Pt, as the
situation is obviously much more complicated than in the single layer case.  The resistivity
of the metallic capping layers is in all our cases lower than that of Py.  Separating the
resistive contributions of capping layers, Py, and interfaces is very difficult, making a
comparison between $\alpha$ and $\rho$ in this case less meaningful. Nonetheless, these
results with adjoining nonmagnetic layers reenforce the above conclusion about the primacy of
electron scattering mechanism in the viscous damping of very thin ferromagnetic films.

The effectiveness of ion-milling in increasing $\alpha$ compared with the lack thereof in the
Cu-coated samples suggests that although the Cu-coated samples are rougher than the very
smooth PR- and Pt-coated samples, the roughness is insignificant as far as magnetization
damping is concerned.  One may expect a much more complex surface after ion-milling, caused
by a mixture of redeposition of Py, surface oxidation and structural defects none of which
are present in the as-deposited films.


In conclusion we have confirmed that magnetization relaxation in ultrathin Py films is
governed by processes that involve ordinary electron scattering.  This is analogous to bulk
relaxation except that the electron scattering in ultrathin films is increasingly caused by
surfaces and defects as the films become thinner, whereas it is to a greater extent caused by
phonons in bulk materials.  We have also observed increased damping in trilayers of Pt/Py/Pt
which we attribute to strong spin-orbit coupling in the Pt-layers.  From a practical
viewpoint our results highlight the important connection between electron scattering and
magnetization damping.  That should prove important when designing magnetic devices with a
desired ``optimal'' dynamic response.

We thank T.~R.~McGuire, L.~Berger, and B.~Heinrich for helpful discussions, and D.~W.~Abraham
for help with AFM and magnetization measurements.  This work was supported in part by
National Science Foundation Grants Nos.\ DMR-0071770 and DMR-0074080.
\bibliography{fmr}

\end{document}